# Schemes for entanglement concentration of two unknown partially entangled states with cross-Kerr nonlinearity

*Wei Xiong, and Liu Ye*

*School of Physics and Material Science, Anhui University, Hefei 230039,*

*People's Republic of China*

**Abstract.** We propose practical schemes for concentrating entanglement of a pair of unknown partially entangled Bell states and three-photon W states with cross-Kerr nonlinearity. In the schemes, utilizing local operations and classical communication, two separated parties can obtain one maximally entangled photon pair from two previously shared partially entangled photon pairs, and three separated parties can obtain one maximally entangled three-photon W state and a maximally entangled cluster state from two identical partially entangled three-photon W state with a certain success probability. Finally, we discuss the influences of sources of errors and de-coherence on the schemes. The proposed setup is very simple, just employing some linear optical elements and cross-Kerr medium, which greatly simplifies the experimental realization of the schemes. The schemes are feasible within current experimental technology,.

## Ⅰ. Introduction

Quantum entanglement between distributed quantum systems is of fundamental importance to the future implementation of various quantum information processing, such as quantum teleportation [1-3], quantum secret key distribution [4-7], quantum computation [8], and quantum dense coding [9,10]. All the above mentioned applications which can be realized ideally with unit success probability and unit fidelity require that two or many distant parties share maximally entangled states. Actually, however, the two or many distant parties can not share maximally entangled states faithfully but some forms of non-maximally entangled pure states can be obtained due to the influences of de-coherence and the imperfection at the sources. Under this condition, the success probability of the implementation will be less than one. For this reason, entanglement concentration [11] schemes are of practical significance because they can extract maximally entangled states from some partially entangled states via applying local operations and classical communication (LOCC). Many theoretical and experimental schemes for obtaining maximally entangled particles by LOCC have been proposed [12-29]. In 1996, Bennett et al. [12] proposed an original entanglement purification scheme for purifying Bell states by use of local operations on copies of noisy Bell pairs and classical communication between two parties. After that, Zhao *et al* [17] proposed a probabilistic scheme for entanglement concentration based on the principle of quantum erasure and the Schmidt projection method. In their scheme, one can concentrate entanglement from arbitrary identical non-maximally entangled pairs at distant locations. Yamamoto et al. [21] proposed an experimentally feasible concentration and purification scheme with linear optical elements such as polarizing beam splitters (PBSs) and quarter wave plates (QWPs). Yang *et al* [22] and Cao *et al* [23] proposed schemes for entanglement concentration of unknown atomic entangled states via entanglement swapping and cavity decay with a certain success probability in cavity QED, respectively.

Although the entanglement concentration in a bipartite system has been studied intensively, there are few schemes for concentrating the non-maximally entangled states of the bipartite system and tripartite system exploiting cross-Kerr nonlinearity. To the best of our knowledge, cross-kerr nonlinearity provides a good tool to construct nondestructive quantum non-demolition (QND) detectors, which have the potential available of being able to condition the evolution of our system but without necessarily destroying the single photons [30, 31]. Such QND detector can determine whether there are photons after the PBS or not, which cannot be accomplished only with PBS. For these reasons, exploiting cross-kerr nonlinearity is full of significances to realize entanglement concentration schemes for a pair of unknown pure non-maximally entangled Bell states and three-photon W states with cross-Kerr nonlinearity in our letter. But interestingly, we not only acquire the maximally entangled Bell states and W states, but also a genuine cluster state, which is essential to the one-way computer [32], can be obtained by LOCC. On the other hand, three-particle W state is easier to prepare than four-particle cluster state. Thus our schemes provide a new way to generate multiple-particle entanglement, which is profound to quantum computer in the further. In the scheme, we use the polarization of photons as qubit and define horizontally (vertically) linear polarization $|H\rangle$ ($|V\rangle$) as the qubit $|0\rangle$ ($|1\rangle$).

## II. Entanglement concentration with weak cross-kerr nonlinearity

Before we outline our schemes of entanglement concentration, we briefly review the principle of QND measurement using weak cross-kerr nonlinearity first presented by Nemoto and Munro [30]. The Hamiltonian of a cross-kerr nonlinear medium can be described by the form as follows (setting $\hbar = 1$):

$$H_{QND} = \chi \hat{n}_a \hat{n}_c \tag{1}$$

where $\hat{n}_a$ ($\hat{n}_c$) denotes the number operator for mode $a$ (c) and $\chi$ is the coupling strength of the nonlinearity, which is decided by the property of material. If we consider a signal state to have the form $|\psi\rangle = a|0\rangle_a + b|1\rangle_a$ with the probe beam initially in a coherent state $|\alpha\rangle_c$, the cross-kerr interaction causes the combined system composed of a single photon and a coherent state to evolve as [30]

$$U_{ck}|\psi\rangle|\alpha\rangle_c = e^{iH_{QND}t}(a|0\rangle_a + b|1\rangle_a)|\alpha\rangle_c = a|0\rangle_a|\alpha\rangle_c + b|1\rangle_a|\alpha e^{i\theta}\rangle_c \tag{2}$$

where $\theta = \chi t$ is introduced by the nonlinearity and t is the interaction time. We observe immediately that the Fock state $|n_a\rangle$ is unaffected by the interaction but the coherent state $|\alpha\rangle_c$ picks up a phase shift directly proportional to the number of photons $n_a$ in the $|n_a\rangle$ state. For $n_a$ photons in the signal mode, the probe beam evolves to $|\alpha e^{in_a\theta}\rangle_c$. Through a general homodyne-heterodyne measurement (X homodyne measurement) of the phase of the coherent state, the signal state $|\psi\rangle$ will be projected into a definite number state or superposition of number states. Because the measurement can be performed with high fidelity, the projection is nearly deterministic. This technique was first used to realize a CNOT gate [30], a parity projector [33], and the Bell state [33]. It provides a new route to new quantum computation [34]. The requirement for this technique is $\alpha\theta \gg 1$ [34], where $\alpha$ is the amplitude of the coherent state. Even with the weak nonlinearity ($\theta$ is small), this requirement can be satisfied with large amplitude of the coherent state. Then this requirement may be feasible with current experimental technology. Our schemes of entanglement concentration also work with the weak cross-kerr nonlinearity.

**A. Concentration scheme for two unknown partially entangled photon pairs**

In this section, we show how the two separated parties Alice and Bob can concentrate a maximally entangled photon pair from two identical partially entangled photon pairs by LOCC. We assume that Alice and Bob are given two pairs of photons in the following polarization entangled states

$$|\psi\rangle_{12}|\psi\rangle_{34} = (\alpha|H\rangle_1|H\rangle_2 + \beta|V\rangle_1|V\rangle_2) \otimes (\alpha|H\rangle_3|H\rangle_4 + \beta|V\rangle_3|V\rangle_4) \qquad (3)$$

where $\alpha$ and $\beta$ are arbitrary complex numbers satisfying $|\alpha|^2 + |\beta|^2 = 1$. Alice holds photons 1 and 3, Bob holds photons 2 and 4, respectively. Alice and Bob can transform these photons into a maximally entangled photon pair in modes $2'$ and $3'$ in the following approach. To begin with, we expand Eq. (3) as

$$|\phi\rangle_1 = \alpha^2|H\rangle_1|H\rangle_2|H\rangle_3|H\rangle_4 + \beta^2|V\rangle_1|V\rangle_2|V\rangle_3|V\rangle_4$$
$$+ \alpha\beta(|H\rangle_1|H\rangle_2|V\rangle_3|V\rangle_4 + |V\rangle_1|V\rangle_2|H\rangle_3|H\rangle_4) \qquad (4)$$

We note that the third term and the fourth term in Eq. (4) have the same coefficients $\alpha\beta$. So we desire to distinguish the third term and the fourth term from the first term and the second term. We let photons in modes 1 and 3 pass through PBS1, PBS 2, PBS 3 and PBS 4. The state of photons is split individually on the PBSs into two spatial modes, which transmit $|H\rangle$ and reflect $|V\rangle$. The probe beam then interacts with the photons in the horizontal mode and vertical mode through the cross-kerr nonlinear medium, the photons on the modes $a_1$ and $a_3$ gets the same phase shift $\theta$ with their Homodyne measurements on their coherent states, as shown in figure 1. So Eq. (4) can be written as

$$|\phi\rangle_2 = \alpha^2 |H\rangle_{a_5} |H\rangle_{a_6} |H\rangle_2 |H\rangle_4 |\alpha e^{i2\theta}\rangle + \beta^2 |V\rangle_{a_5} |V\rangle_{a_6} |V\rangle_2 |V\rangle_4$$
$$+ \alpha\beta(|H\rangle_{a_5} |V\rangle_{a_6} |H\rangle_2 |V\rangle_4 + |V\rangle_{a_5} |H\rangle_{a_6} |V\rangle_2 |H\rangle_4) |\alpha e^{i\theta}\rangle \qquad (5)$$

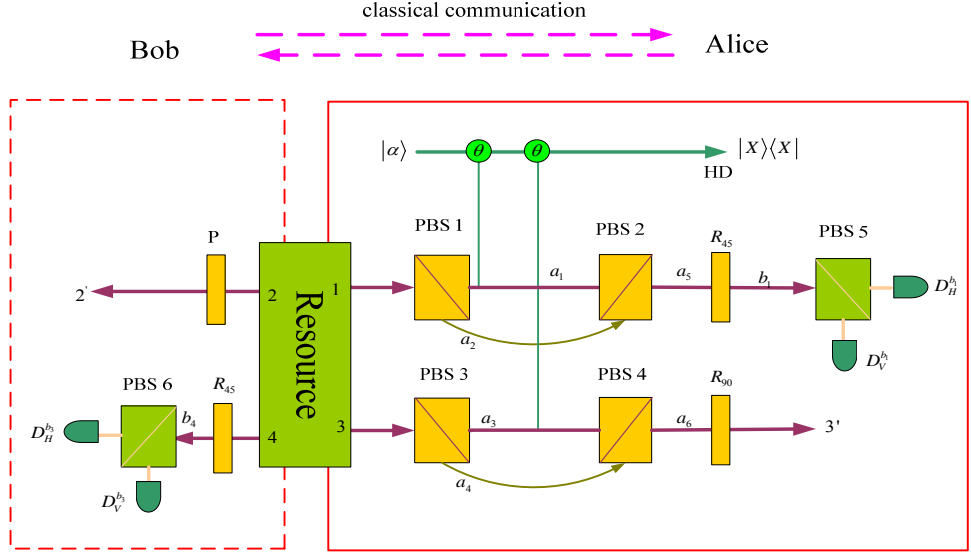

Figure 1. (Color online) The schematic diagram of the proposed entanglement concentration for two unknown partially entangled photon pairs. Polarization beam splitters(PBS) transmit H photons and reflect V photons. HD denotes homodyne measurement, R denotes $\lambda/2$ wave plate, P represents $\pi/2$-phase shifter, and D denotes detector.

Through a general X homodyne measurement, if we find that probe mode is in the coherent state $|\alpha e^{i\theta}\rangle$, the four-photon state will be projected into the following state:

$$|\phi\rangle_3 = \alpha\beta(|H\rangle_{a_5} |V\rangle_{a_6} |H\rangle_2 |V\rangle_4 + |V\rangle_{a_5} |H\rangle_{a_6} |V\rangle_2 |H\rangle_4) \qquad (6)$$

In the following, Alice and Bob rotate the polarizations of their photons in modes 4, $a_5$ and $a_6$ by $45°$ and $90°$ using $\lambda/2$ wave plates ($R_{45}$ and $R_{90}$), the action of $R_{45}$ is given by

$$|H\rangle \to \frac{1}{\sqrt{2}}(|H\rangle+|V\rangle), \quad |V\rangle \to \frac{1}{\sqrt{2}}(|H\rangle-|V\rangle) \quad (7)$$

and the action of $R_{90}$ makes $|H\rangle \leftrightarrow |V\rangle$. Therefore, Eq.(6) is changed into the state

$$\begin{aligned}|\phi\rangle_4 = \frac{1}{2}(&|H\rangle|H\rangle|H\rangle|H\rangle - |H\rangle|V\rangle|H\rangle|H\rangle \\
&+|V\rangle|H\rangle|H\rangle|H\rangle - |V\rangle|V\rangle|H\rangle|H\rangle \\
&+|H\rangle|H\rangle|V\rangle|V\rangle + |H\rangle|V\rangle|V\rangle|V\rangle \\
&-|V\rangle|H\rangle|V\rangle|V\rangle - |V\rangle|V\rangle|V\rangle|V\rangle)_{b_1 b_4 3'2'} \end{aligned} \quad (8)$$

Finally, let the photons in modes $b_1$ and $b_4$ pass through PBS 5 and PBS 6, respectively. Apparently, if Alice and Bob detect the photons in the polarization state $|H\rangle_{b_1}|H\rangle_{b_4}$ ($|V\rangle_{b_1}|V\rangle_{b_4}$), then the remaining two photons in modes 3' and 2' are left in the state

$$|\phi\rangle^+ = \frac{1}{\sqrt{2}}(|H\rangle|H\rangle + |V\rangle|V\rangle)_{3'2'} \quad (9)$$

Similarly, if Alice and Bob detect the photons in the polarization state $|H\rangle_{b_1}|V\rangle_{b_4}$ ($|V\rangle_{b_1}|H\rangle_{b_4}$), then the remaining two photons in modes 3' and 2' are left in the state

$$|\phi\rangle^- = \frac{1}{\sqrt{2}}(|H\rangle|H\rangle - |V\rangle|V\rangle)_{3'2'} \quad (10)$$

which can be transformed into equation (9) by applying a $\pi/2$-phase shifter P to change the sign of the polarization state $|V\rangle_2$. Therefore, the total probability of sharing a maximally entangled photon pair in the state $|\phi\rangle^+$ is $2|\alpha|^2(1-|\alpha|^2)$, which is plotted in figure 4(a).

**B. Concentration scheme for two partially entangled three-photon W state**

In the following, we assume that three parties Alice, Bob and Charlie share an unknown three-photon polarization entangled states:

$$|\varphi\rangle_{456} = \gamma|H\rangle_4|H\rangle_5|V\rangle_6 + \delta(|H\rangle_4|V\rangle_5|H\rangle_6 + |V\rangle_4|H\rangle_5|H\rangle_6) \quad (11)$$

where $\gamma$ and $\delta$ are arbitrary complex numbers satisfying $|\gamma|^2 + 2|\delta|^2 = 1$. Alice holds photons 6, Bob and Charlie hold photons 4 and 5, respectively. Furthermore, Alice also holds an identical state (11) on her hand $|\varphi\rangle_{123}$. Therefore, the state of the whole system is given by

$$|\varphi\rangle_{123} \otimes |\varphi\rangle_{456} = [\gamma |H\rangle_1 |H\rangle_2 |V\rangle_3 + \delta(|H\rangle_1 |V\rangle_2 |H\rangle_3 + |V\rangle_1 |H\rangle_2 |H\rangle_3)]$$

$$\otimes [\gamma |H\rangle_4 |H\rangle_5 |V\rangle_6 + \delta(|H\rangle_4 |V\rangle_5 |H\rangle_6 + |V\rangle_4 |H\rangle_5 |H\rangle_6)_{456}] \quad (12)$$

Alice, Bob and Charlie can transform these photons into a maximally entangled three-photon W state in modes $2'$, $4'$ and $5'$ in the following way. First, we expand Eq. (12) as

$$|\Phi\rangle = \gamma^2 |H\rangle_1 |H\rangle_2 |V\rangle_3 |H\rangle_4 |H\rangle_5 |V\rangle_6$$

$$+ \gamma\delta(|H\rangle_1 |H\rangle_2 |V\rangle_3 |H\rangle_4 |V\rangle_5 |H\rangle_6 + |H\rangle_1 |H\rangle_2 |V\rangle_3 |V\rangle_4 |H\rangle_5 |H\rangle_6$$

$$+ |H\rangle_1 |V\rangle_2 |H\rangle_3 |H\rangle_4 |H\rangle_5 |V\rangle_6 + |V\rangle_1 |H\rangle_2 |H\rangle_3 |H\rangle_4 |H\rangle_5 |V\rangle_6)$$

$$+ \delta^2(|H\rangle_1 |V\rangle_2 |H\rangle_3 |H\rangle_4 |V\rangle_5 |H\rangle_6 + |H\rangle_1 |V\rangle_2 |H\rangle_3 |V\rangle_4 |H\rangle_5 |H\rangle_6$$

$$+ |V\rangle_1 |H\rangle_2 |H\rangle_3 |H\rangle_4 |V\rangle_5 |H\rangle_6 + |V\rangle_1 |H\rangle_2 |H\rangle_3 |V\rangle_4 |H\rangle_5 |H\rangle_6) \quad (13)$$

We observe that there are four terms which have the same coefficient $\gamma\delta$. Therefore, we want to distinguish them from the state (13). We send photons 1, 2, 3 and 6 to pass through PBS j(j=1,2……8), then the probe beam interacts with the photons in the horizontal mode and vertical mode through the cross-kerr nonlinear medium, the photons in modes $a_1$ and $a_3$ get the same phase shift $\theta$, the photon in mode $a_5$ gets a phase shift $-\theta$ and the photon in mode $a_6$ picks up a phase shift $-2\theta$ with their x Homodyne measurements on their coherent states, as shown in figure 2. Consequently, equation (13) can be written as

$$|\Phi\rangle_1 = \gamma^2 |H\rangle_{a_1} |H\rangle_{a_3} |V\rangle_{a_6} |V\rangle_{a_8} |H\rangle_4 |H\rangle_5 |\alpha e^{i2\theta}\rangle$$

$$+ \gamma\delta(|H\rangle_{a_1} |H\rangle_{a_3} |V\rangle_{a_6} |H\rangle_{a_7} |H\rangle_4 |V\rangle_5 + |H\rangle_{a_1} |H\rangle_{a_3} |V\rangle_{a_6} |H\rangle_{a_7} |V\rangle_4 |H\rangle_5$$

$$+ |H\rangle_{a_1} |V\rangle_{a_4} |H\rangle_{a_5} |V\rangle_{a_8} |H\rangle_4 |H\rangle_5 + |V\rangle_{a_2} |H\rangle_{a_3} |H\rangle_{a_5} |V\rangle_{a_8} |H\rangle_4 |H\rangle_5)|\alpha\rangle$$

$$+\delta^2(|H\rangle_{a_1}|V\rangle_{a_4}|H\rangle_{a_5}|H\rangle_{a_7}|H\rangle_4|V\rangle_5+|H\rangle_{a_1}|V\rangle_{a_4}|H\rangle_{a_5}|H\rangle_{a_7}|V\rangle_4|H\rangle_5$$

$$+|V\rangle_{a_2}|H\rangle_{a_3}|H\rangle_{a_5}|H\rangle_{a_7}|H\rangle_4|V\rangle_5+|V\rangle_{a_2}|H\rangle_{a_3}|H\rangle_{a_5}|H\rangle_{a_7}|V\rangle_4|H\rangle_5)|\alpha e^{-i2\theta}\rangle \quad (14)$$

If we find that probe mode is in the coherent state $|\alpha\rangle$, the above state will be projected to the state:

$$|\Phi\rangle_2 = \gamma\delta(|H\rangle_{a_9}|H\rangle_{a_{10}}|V\rangle_{a_{11}}|H\rangle_{a_{12}}|H\rangle_4|V\rangle_5+|H\rangle_{a_9}|H\rangle_{a_{10}}|V\rangle_{a_{11}}|H\rangle_{a_{12}}|V\rangle_4|H\rangle_5$$

$$+|H\rangle_{a_9}|V\rangle_{a_{10}}|H\rangle_{a_{11}}|V\rangle_{a_{12}}|H\rangle_4|H\rangle_5+|V\rangle_{a_9}|H\rangle_{a_{10}}|H\rangle_{a_{11}}|V\rangle_{a_{12}}|H\rangle_4|H\rangle_5) \quad (15)$$

In the following, to begin with, Alice rotates the polarizations of her photons in mode $a_{11}$ by 90° using $\lambda/2$ wave plates ($R_{90}$), then she rotates the polarizations of her photons in mode $a_{11}$ by 45° using $\lambda/2$ wave plates ($R_{45}$), respectively, as shown in figure 2. So equation (15) will be given by

$$|\Phi\rangle_3 = \frac{\gamma\delta}{\sqrt{2}}[(|H\rangle+|V\rangle)_{a_{9'}}|H\rangle_{a_{10}}|H\rangle_{a_{11'}}|H\rangle_{a_{12}}|H\rangle_4|V\rangle_5+(|H\rangle+|V\rangle)_{a_{9'}}|H\rangle_{a_{10}}|H\rangle_{a_{11'}}|H\rangle_{a_{12}}|V\rangle_4|H\rangle_5$$

$$+(|H\rangle+|V\rangle)_{a_{9'}}|V\rangle_{a_{10}}|V\rangle_{a_{11'}}|V\rangle_{a_{12}}|H\rangle_4|H\rangle_5+(|H\rangle-|V\rangle)_{a_{9'}}|H\rangle_{a_{10}}|V\rangle_{a_{11'}}|V\rangle_{a_{12}}|H\rangle_4|H\rangle_5]$$

$$(16)$$

Then we send photons in modes $a_{9'}$, $a_{10}$, $a_{11'}$ and $a_{12}$ to pass through PBS i ( i =9,10,……16), after that, the probe beam interacts with the photons in the horizontal mode and vertical mode through the cross-kerr nonlinear medium, and we find the photons in modes $a_{14}$ and $a_{15}$ get the same phase shift $\theta$, the photon in modes $a_{17}$ and $a_{20}$ pick up the same phase shift $-\theta$. Therefore, Eq.(16) evolves to the state

$$|\Phi\rangle_4 = \frac{\gamma\delta}{\sqrt{2}}[(|H\rangle_{1'}|H\rangle_{2'}|H\rangle_{3'}|H\rangle_{6'}|H\rangle_4|V\rangle_5+|H\rangle_{1'}|H\rangle_{2'}|H\rangle_{3'}|H\rangle_{6'}|V\rangle_4|H\rangle_5$$

$$+|V\rangle_{1'}|V\rangle_{2'}|V\rangle_{3'}|V\rangle_{6'}|H\rangle_4|H\rangle_5)|\alpha\rangle+(|V\rangle_{1'}|H\rangle_{2'}|H\rangle_{3'}|H\rangle_{6'}|V\rangle_4|H\rangle_5$$

$$+|V\rangle_{1'}|H\rangle_{2'}|H\rangle_{3'}|H\rangle_{6'}|H\rangle_4|V\rangle_5)|\alpha e^{-i\theta}\rangle+(|H\rangle_{1'}|V\rangle_{2'}|V\rangle_{3'}|V\rangle_{6'}|H\rangle_4|H\rangle_5$$

$$-|V\rangle_{1'}|H\rangle_{2'}|V\rangle_{3'}|V\rangle_{6'}|H\rangle_4|H\rangle_5)|\alpha e^{i\theta}\rangle+|H\rangle_{1'}|H\rangle_{2'}|V\rangle_{3'}|V\rangle_{6'}|H\rangle_4|H\rangle_5|\alpha e^{i2\theta}\rangle](17)$$

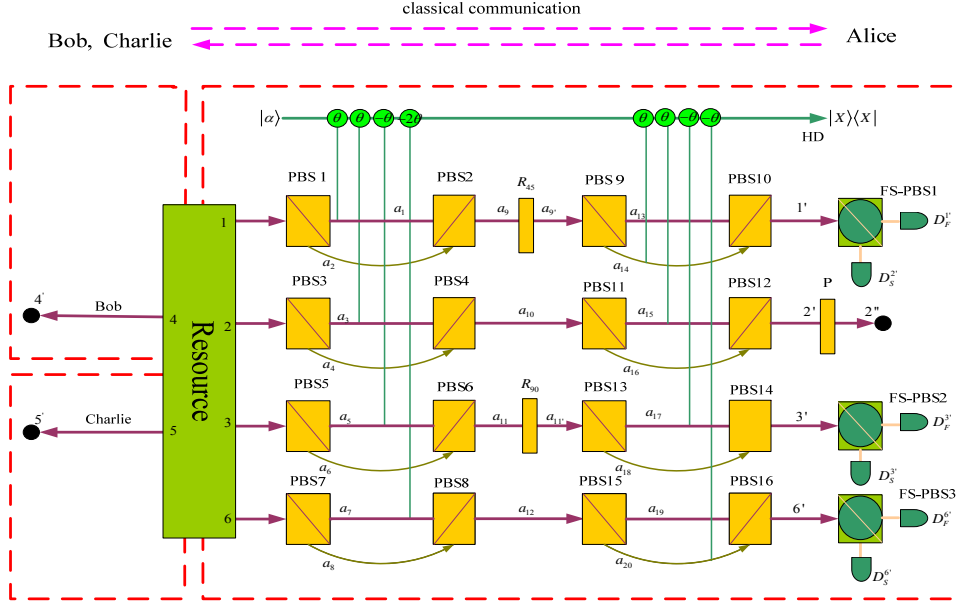

Figure 2. (Color online) The schematic diagram of the proposed entanglement concentration for two unknown partially entangled three-photon W state. Rotated Polarization beam splitters(FS-PBS) transmit F polarization photons and reflect S polarization photons. HD denotes homodyne measurement, R denotes $\lambda/2$ wave plate, P represents $\pi/2$-phase shifter, and D denotes detector.

Through a general X homodyne measurement, we find if the probe beam is in the coherent state $|\alpha\rangle$, the state (17) will be projected to the state

$$|\Phi\rangle_5 = \frac{\gamma\delta}{\sqrt{2}}(|H\rangle_{1'}|H\rangle_{2'}|H\rangle_{3'}|H\rangle_{6'}|H\rangle_4|V\rangle_5 + |H\rangle_{1'}|H\rangle_{2'}|H\rangle_{3'}|H\rangle_{6'}|V\rangle_4|H\rangle_5$$
$$+ |V\rangle_{1'}|V\rangle_{2'}|V\rangle_{3'}|V\rangle_{6'}|H\rangle_4|H\rangle_5) \qquad (18)$$

Finally, let the photons in modes $1'$, $3'$ and $6'$ pass through a series of rotated polarizing beam splitters (FS-PBS) k (k=1, 2, 3), which change $|V\rangle$ and $|H\rangle$ into a new frame as

$$|V\rangle = \frac{1}{\sqrt{2}}(|F\rangle + |S\rangle), \quad |H\rangle = \frac{1}{\sqrt{2}}(|F\rangle - |S\rangle) \qquad (19)$$

and always reflect S-polarizing photons and transmit F-polarizing photons. In the new frame, the state given in Eq. (18) can become into

$$|\Phi\rangle_6 = \frac{\gamma\delta}{4}[|F\rangle_{1'}|F\rangle_{3'}|F\rangle_{6'}(|H\rangle|H\rangle|V\rangle + |H\rangle|V\rangle|H\rangle + |V\rangle|H\rangle|H\rangle)_{2'4'6'}$$

$$+|F\rangle_{1'}|S\rangle_{3'}|S\rangle_{6'}(|H\rangle|H\rangle|V\rangle+|H\rangle|V\rangle|H\rangle+|V\rangle|H\rangle|H\rangle)_{2'4'6'}$$

$$+|S\rangle_{1'}|S\rangle_{3'}|F\rangle_{6'}(|H\rangle|H\rangle|V\rangle+|H\rangle|V\rangle|H\rangle+|V\rangle|H\rangle|H\rangle)_{2'4'6'}$$

$$+|S\rangle_{1'}|F\rangle_{3'}|S\rangle_{6'}(|H\rangle|H\rangle|V\rangle+|H\rangle|V\rangle|H\rangle+|V\rangle|H\rangle|H\rangle)_{2'4'6'}$$

$$-|F\rangle_{1'}|F\rangle_{3'}|S\rangle_{6'}(|H\rangle|H\rangle|V\rangle+|H\rangle|V\rangle|H\rangle-|V\rangle|H\rangle|H\rangle)_{2'4'6'}$$

$$-|F\rangle_{1'}|S\rangle_{3'}|F\rangle_{6'}(|H\rangle|H\rangle|V\rangle+|H\rangle|V\rangle|H\rangle-|V\rangle|H\rangle|H\rangle)_{2'4'6'}$$

$$-|S\rangle_{1'}|F\rangle_{3'}|F\rangle_{6'}(|H\rangle|H\rangle|V\rangle+|H\rangle|V\rangle|H\rangle-|V\rangle|H\rangle|H\rangle)_{2'4'6'}$$

$$-|S\rangle_{1'}|S\rangle_{3'}|S\rangle_{6'}(|H\rangle|H\rangle|V\rangle+|H\rangle|V\rangle|H\rangle-|V\rangle|H\rangle|H\rangle)_{2'4'6'}] \qquad (20)$$

Obviously, when Alice detects the photons in the polarization state $|F\rangle_{1'}|F\rangle_{3'}|F\rangle_{6'}$ ($|F\rangle_{1'}|S\rangle_{3'}|S\rangle_{6'}$, $|S\rangle_{1'}|S\rangle_{3'}|F\rangle_{6'}$ or $|S\rangle_{1'}|F\rangle_{3'}|S\rangle_{6'}$), equation (20) will be projected into the state

$$|W^+\rangle = \frac{1}{\sqrt{3}}(|H\rangle|H\rangle|V\rangle+|H\rangle|V\rangle|H\rangle+|V\rangle|H\rangle|H\rangle)_{2'4'6'} \qquad (21)$$

Similarly, if Alice detects the photons in the polarization state $|S\rangle_{1'}|S\rangle_{3'}|S\rangle_{6'}$ ($|F\rangle_{1'}|F\rangle_{3'}|S\rangle_{6'}$, $|F\rangle_{1'}|S\rangle_{3'}|F\rangle_{6'}$, $|S\rangle_{1'}|F\rangle_{3'}|F\rangle_{6'}$), equation (20) will be projected into the state

$$|W^-\rangle = \frac{1}{\sqrt{3}}(|H\rangle|H\rangle|V\rangle+|H\rangle|V\rangle|H\rangle-|V\rangle|H\rangle|H\rangle)_{2'4'6'} \qquad (22)$$

which can be transformed into equation (21) by utilizing a $\pi/2$-phase shifter P to change the sign of the polarization state $|V\rangle_{2'}$ into $-|V\rangle_{2''}$. Therefore, the total success probability of obtaining three-photon W state is $\frac{3}{2}|\gamma\delta|^2$, which is plotted in figure 4(a).

Compared to [35], the probability of our scheme for concentrating a maximally entangled three-photon W state is about $9/2$ times than theirs, and our scheme is easier to operate in practical realization, which is feasible within current technology.

### C. Extracting a maximally entangled four-photon cluster state from two unknown partially entangled three-photon W state

From equation (14), we find that if probe mode is in the coherent state $|\alpha e^{-i2\theta}\rangle$ through X homodyne measurement, the state in equation (14) will be projected to the state:

$$|\Psi\rangle_1 = \delta^2(|H\rangle_{b_1}|V\rangle_{b_2}|H\rangle_{b_3}|H\rangle_{b_4}|V\rangle_{b_5}|H\rangle_{b_6} + |H\rangle_{b_1}|V\rangle_{b_2}|H\rangle_{b_3}|V\rangle_{b_4}|H\rangle_{b_5}|H\rangle_{b_6}$$
$$+ |V\rangle_{b_1}|H\rangle_{b_2}|H\rangle_{b_3}|H\rangle_{b_4}|V\rangle_{b_5}|H\rangle_{b_6} + |V\rangle_{b_1}|H\rangle_{b_2}|H\rangle_{b_3}|V\rangle_{b_4}|H\rangle_{b_5}|H\rangle_{b_6}) \quad (23)$$

The setup is shown in figure 3. Subsequently, Alice and Charile send photons in modes $b_2$ and $b_5$ to pass through $\pi$ cross-kerr medium. The action of $\pi$-cross-Kerr medium evolves the modes $b_2$ and $b_5$ as follows from Eq. (1), which in the polarization basis produces a $\pi$ phase shift on the $|VV\rangle$ term. Conceptually, the simplest such two-qubit gate is the CZ gate, i.e.,

$$|HH\rangle \to |HH\rangle; \quad |HV\rangle \to |HV\rangle; \quad |VH\rangle \to |VH\rangle; \quad |VV\rangle \to -|VV\rangle. \quad (24)$$

After that, Alice and Charlie rotate the polarizations of her photons in mode $b_1$ and $b_4$ by $90°$ using $\lambda/2$ wave plates ($R_{90}$), as shown in figure 3. Thus Eq. (23) will evolve to the state

$$|\Psi\rangle_2 = \delta^2(-|V\rangle_{b_{1'}}|V\rangle_{b_2}|H\rangle_{b_3}|V\rangle_{b_{4'}}|V\rangle_{b_5}|H\rangle_{b_6} + |V\rangle_{b_{1'}}|V\rangle_{b_2}|H\rangle_{b_3}|H\rangle_{b_{4'}}|H\rangle_{b_5}|H\rangle_{b_6}$$
$$+ |H\rangle_{b_{1'}}|H\rangle_{b_2}|H\rangle_{b_3}|V\rangle_{b_{4'}}|V\rangle_{b_5}|H\rangle_{b_6} + |H\rangle_{b_{1'}}|H\rangle_{b_2}|H\rangle_{b_3}|H\rangle_{b_{4'}}|H\rangle_{b_5}|H\rangle_{b_6}) \quad (25)$$

Finally, Alice detects the photons in modes $b_3$ and $b_6$, and we find Alice's result is $|H\rangle_{b_3}|H\rangle_{b_6}$. Following this way, Alice, Bob and Charlie share a maximally entangled cluster state

$$|C\rangle = \frac{1}{2}(|H\rangle_{b_{1'}}|H\rangle_{b_2}|H\rangle_{b_{4'}}|H\rangle_{b_5} + |V\rangle_{b_{1'}}|V\rangle_{b_2}|H\rangle_{b_{4'}}|H\rangle_{b_5}$$
$$+ |H\rangle_{b_{1'}}|H\rangle_{b_2}|V\rangle_{b_{4'}}|V\rangle_{b_5} - |V\rangle_{b_{1'}}|V\rangle_{b_2}|V\rangle_{b_{4'}}|V\rangle_{b_5}) \quad (26)$$

The success probability is $4|\delta|^4$, which is also plotted in figure 4(a).

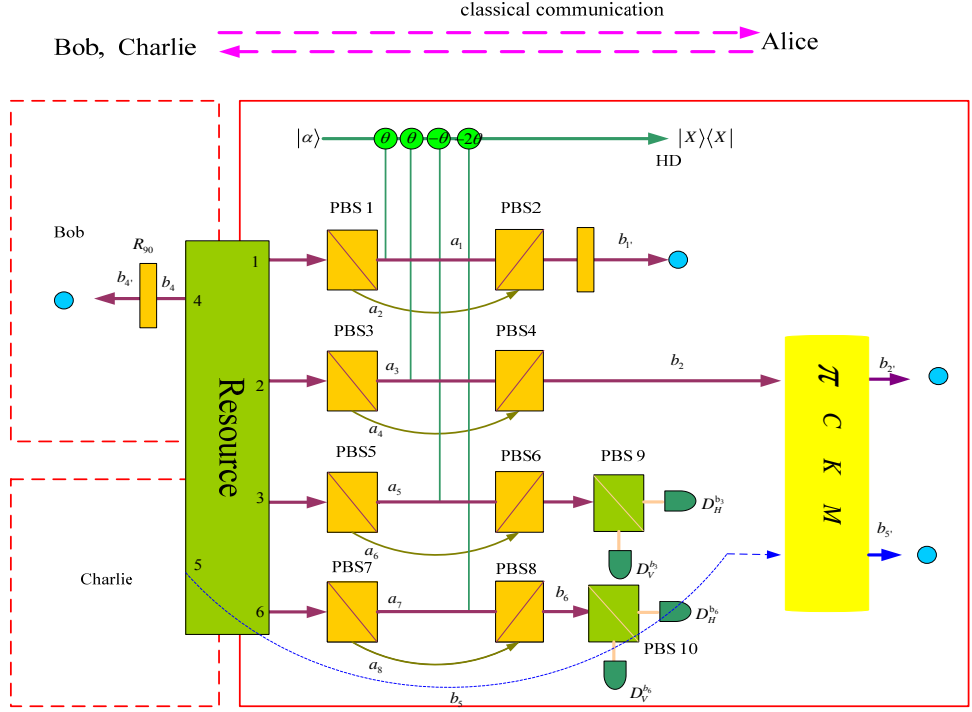

Figure 3. (Color online) The schematic diagram of the proposed entanglement concentration for two unknown partially entangled four-photon cluster state. $\pi$ CKM: $\pi$ cross-Kerr medium.

## III. Discussion and Analysis

In this section, let us briefly analyze and discuss some practical issues in relation to the experimental feasibility of our schemes. Firstly, we simply discuss the sources of errors and their effects. There are two types of errors on the probe mode: (1) an intrinsic measurement error which arises from the fact that phase-shifted coherent states $|\alpha e^{\pm i\theta}\rangle$ and coherent state $|\alpha\rangle$ of the probe mode are not completely orthogonal and a measurement result in one parity subspace could have come form the opposite parity state. This intrinsic error, given by $P_{err}(\theta) = erfc[|\alpha|\sin\theta/\sqrt{2}]/2$ which can be suppressed (made small) when $|\alpha\theta| \gg 1$ [30]. For instance with $\alpha\theta \sim \pi$, $P_{err} \sim 10^{-3}$. Choosing the mean photon number per pulse to be on the order of $10^{12}$ (corresponding to $\alpha \sim 10^6$) in the realistic pumps, a weak nonlinearity $\theta \sim 3.14 \times 10^{-6}$ should be sufficient to satisfy $\alpha\theta \sim \pi$. (2) Errors due to photon loss, decoherence or phase noise on the probe mode. In the real situation, decoherence is inevitable, the photon loss may occur when the coherent state transmitted through an optical fiber. When photon loss occurs, the qubit states will evolve to the mixed states after the homodyne detection [36-38], after which the fidelity of the proposed schemes will degrade. As described above, the amplitude of the coherent state $\alpha$ may be large enough to satisfy the requirement $\alpha\theta > 1$ when the cross-Kerr nonlinearity is small. However, as the increasing of the amplitude of the coherent states, the fidelity of these schemes will decrease simultaneously due to the decoherence (photon loss). Fortunately, the decoherence can be made arbitrarily small simply by an arbitrary strong coherent state associated with a displacement $D(-\alpha)$ performed on the coherent state and the QND photon-number-resolving detection [37]. Additionally, the photon loss also causes de-phasing, corresponding to phase flip errors, in the original two-qubit state [36]. The degree of de-phasing is characterized by the parameters $\gamma = \eta^2\alpha^2\theta^2/2$ where $\eta^2$ is the percentage of photons lost from the probe mode. We plot it to analyze more distinctly in figure 4(b). $\gamma$ must be keep small for the de-phasing to have a negligible effect. This in effect requires $\eta \ll 1/|\alpha|\theta$ which can be simply satisfied as long $1 \ll |\alpha|\theta < 10$. For instance with $\alpha\theta \sim \pi$ and $\eta \sim 0.035$. The error due to de-phasing is of the order $10^{-3}$. Now as we make $|\alpha|\theta$ larger, η needs to decrease to keep the de-phasing error small. Other potential errors worth mentioning are associated with differences in θ between the various qubits and uncertainty in the value of θ. Both of these errors can be managed and are small when $\Delta\theta/\theta \ll 1$.

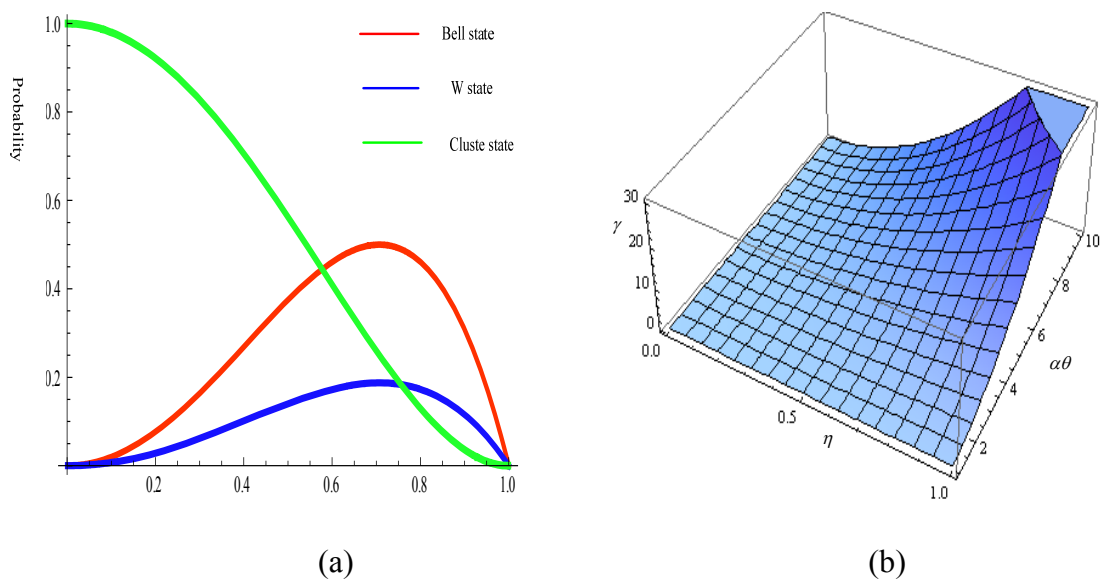

(a)                        (b)

Figure 4: (Color on line) (a) The probabilities of concentrating Bell state (red curve), W state (blue curve) and cluster state(green curve). (b) The degree of de-phasing $\gamma$ as the function of $\eta$ and $\alpha\theta$.

Secondly, we analyze some self-kerr nonlinearity introduced by cross-kerr nonlinearity to our schemes. Self-kerr nonlinearity will cause self-phase modulation (SPM). In the continuous-time framework [39], the phase noises, due to SPM and dispersion etc., will be seen to severely degrade the fidelity of the proposed schemes. (However, Matsuda et al. [40] found that free-carrier dispersion as well as the optical kerr effect contributes to the XPM.) It will also be shown that the phase noise is proportional to the response function's amplitude, implying that stronger nonlinearity is accompanied by increased phase noise. In the above analysis, we have assumed that there is no SPM in our nonlinear material. It turns out that even if the SPM effect is present in the medium, it can be suppressed by operating in the slow-response regime. In Ref. [41], one scheme for the avoidance of the self-modulation effect is to use a resonant $\chi^{(3)}$ medium. The measurement accuracy and the imposed phase noise on the signal wave satisfy the Heisenberg uncertainty principle. This demonstrates that the minimum product of the measurement accuracy and the imposed uncertainty on the conjugate observable is achievable in the proposed QND measurement. Therefore, we only consider the ideal conditions, and the phase noise is not taken into account in our schemes.

Furthermore, schemes need not use collective measurement by photon detectors and need not the states to be known beforehand, so we can use the conventional photon detectors that can only distinguish the vacuum and non-vacuum Fock number states instead of using the sophisticated single-photon detectors distinguishing one or two photon states. The total success probabilities of our schemes are $P_{EPR} = 2|\mu\alpha\beta|^2$, $P_{w} = \frac{3}{2}|\gamma\delta|^2 \mu^3$ and $P_{cluster} = 4|\delta|^4 \mu^2$, with $\mu$ being the quantum efficiency. For scalable linear optics quantum computation, the required quantum efficiency $\mu$ of the

single-photon detectors is extremely high, e.g. for gate success with probability $P \simeq 0.99$, $\mu \gg 0.999987$ [42]. Although experiments for single-photon detectors have made tremendous progress, such detectors still go beyond the current experimental technologies. This greatly decreases the high-quality requirements of photon detectors in practical realization. For the giant cross-kerr nonlinearities, i.e. $\pi$, Ref. [43, 44] has shown that the nonlinear interaction between weak optical pulses can be dramatically enhanced by a technique for generating stationary pulses, where a $\pi$ phase shift is achievable. The technique used in Ref. [44] has been experimentally demonstrated [45]. Therefore, our schemes are experimentally feasible and can be implemented.

### Ⅳ. Conclusion

We have proposed experimentally feasible schemes to realize entanglement concentration of two unknown partially entangled states with cross-kerr nonlinearity. In the schemes, separated parties can ontain a maximally entangled photon state from two identical partially entangled photon states by LOCC. The effects of sources of errors and de-coherence have been investigated. We have shown that most effects of de-coherence can be suppressed (or negligible) under current experimental technologies. Additionally, just employing some linear optical elements and cross-kerr medium, the schemes are feasible in current experimental technology. Following the technology developed in experiments on multi-photon entanglement engineering [46-49] and quantum communication [2], our schemes may be useful for long-distance quantum information processing and quantum communication in the future.

This work was supported by the National Science Foundation of China under Grant No. 11074002, the Doctoral Foundation of the Ministry of Education of China under Grant No. 20103401110003, and also by the Personal Development Foundation of Anhui Province (2008Z018).